# Clustering to Maximize the Ratio of Split to Diameter

**Jiabing Wang**                                                           JBWANG@SCUT.EDU.CN
**Jiaye Chen**                                                        CHEN.JIAYE@MAIL.SCUT.EDU.CN
School of Computer Science and Engineering, South China University of Technology, Guangzhou 510006, China

## Abstract

Given a weighted and complete graph $G = (V, E)$, $V$ denotes the set of $n$ objects to be clustered, and the weight $d(u, v)$ associated with an edge $(u, v) \in E$ denotes the dissimilarity between objects $u$ and $v$. The *diameter* of a cluster is the maximum dissimilarity between pairs of objects in the cluster, and the *split* of a cluster is the minimum dissimilarity between objects within the cluster and objects outside the cluster. In this paper, we propose a new criterion for measuring the goodness of clusters—the ratio of the minimum split to the maximum diameter, and the objective is to maximize the ratio. For $k = 2$, we present an exact algorithm. For $k \geq 3$, we prove that the problem is NP-hard and present a factor of 2 approximation algorithm on the precondition that the weights associated with $E$ satisfy the triangle inequality. The worst-case runtime of both algorithms is $O(n^3)$. We compare the proposed algorithms with the Normalized Cut by applying them to image segmentation. The experimental results on both natural and synthetic images demonstrate the effectiveness of the proposed algorithms.

## 1. Introduction

Clustering groups a set of objects in a way that minimizes the intra-cluster dissimilarity and maximizes the inter-cluster dissimilarity. An ideal cluster can be defined as a set of objects that is compact and isolated [Jain, 2010]: if a cluster is compact then the cluster satisfies the homogeneity criterion of a cluster, and if a cluster is isolated from other clusters, then the cluster satisfies the separation criterion of a cluster. Many clustering algorithms consider only separation or only homogeneity criterion, e.g., the well-known single-linkage clustering and complete-linkage clustering, the former maximizes the minimum dissimilarity between different clusters, and the later attempts to minimize the maximum dissimilarity within the same cluster.

The *diameter* of a cluster is the maximum dissimilarity between pairs of objects within the same cluster, and the *split* of a cluster is the minimum dissimilarity between objects within the cluster and objects outside the cluster. Clearly, the diameter of a cluster is a natural indication of homogeneity of the cluster and the split of a cluster is a natural indication of separation between the cluster and other clusters. Therefore, many clustering algorithms have been proposed for minimizing the maximum diameter or radii of clusters (*minmax* diameter problem), or maximizing the minimum split of clusters (*maxmin* split problem).

Gonzalez proved that the *minmax* diameter problem is NP-hard and a simple 2-approximation algorithm was proposed in [Gonzalez, 1985]. At the same time, Gonzalez also showed that one cannot approximate the optimal solution within an approximation ratio close to 2 in polynomial time unless P = NP. Feder and Greene [Feder & Greene, 1988] also shows that it is NP-hard to approximate the Euclidean *minmax* radius $k$-clustering with an approximation ratio smaller than 1.822, or the Euclidean *minmax* diameter $k$-clustering with an approximation ratio smaller than 1.969, where $k$ is the number of clusters. At the same time, they also proposed an $O(n\log k)$ algorithm for *minmax* diameter or radius problem, where $n$ is the number of objects. Whereas the *minmax* diameter problem is NP-hard for $k \geq 3$, the *maxmin* split problem can be solved using the single-linkage clustering for any $k$ [Delattre & Hansen, 1980].

Then, an ideal partition should have a smaller diameter and a larger split. However, the two criteria are often conflicting. The *minmax* diameter clustering often suffers from the *dissection effect* [Cormack, 1971]. On the other hand, the *maxmin* split clustering suffers from the *chain effect* [Johnson, 1967]. Therefore, a combination of homogeneity and separation conditions may conduce to overcome the drawbacks resulted from a single condition. A number of criteria have been proposed in order to achieve the goal. However, considering the space limit, it is impossible to give a detailed review about those works, as well as include them in the references.

In this paper, we study the following optimization problem: maximize the ratio of the minimum split to the

---

maximum diameter. This combinatory clustering criterion seems very natural and captures both the homogeneity and the separation conditions which a better clustering algorithm should be satisfied.

The rest of this paper is organized as follows. Section 2 introduces some concepts of graph theory relevant to this work and the problem formulation. We prove that the problem is NP-hard in Section 3. Section 4 presents an exact algorithm for $k = 2$ and a 2-approximation algorithm for $k \geq 3$ along with the complexity analysis. Section 5 presents the experimental results. We conclude the paper in Section 6.

## 2. Preliminary and Problem Formulation

We recall several concepts of graph theory [West, 2001]. An undirected graph $G = (V, E)$ consists of a set $V$ of *vertices* and a set $E$ of pairs of vertices called *edges*. For a graph $G = (V, E)$, the *complementary graph* $G^- = (V, E^-)$ of $G$ is a graph with the same set $V$ of vertices as $G$ and with an edge $(u, v) \in E^-$ if and only if $(u, v) \notin E$. A graph $G = (V, E)$ is *complete* if for each pair of vertices $u$ and $v$ of $V$, $(u, v) \in E$. A set of vertices $A \subseteq V$ is a *clique* if and only if every two vertices of $A$ are joined by an edge: $\forall u, v \in A, (u, v) \in E$.

*A coloring* of a graph is a labeling of vertices where adjacent vertices do not share a label. The labels are then often called *colors*. The vertices of a graph are *k-colorable*, or simply a graph is *k-colorable*, if the vertices of a graph can be colored using (at most) $k$ colors. The smallest number $k$ such that a graph $G$ is $k$-colorable, is called the *chromatic number* of $G$, denoted by $\chi(G)$. Given a graph $G$, the problem of whether $G$ is $k$-colorable is NP-complete and the decision of $\chi(G)$ is NP-hard for $k \geq 3$ [Garey & Johnson, 1979]. For $k = 2$, the problem is solvable in polynomial time [Cormen et al, 2001; West, 2001]. Note that a tree is always 2-colorable and the following is a bicoloring algorithm for it: select an arbitrary vertex and color it *black*, and suppose $S$ is the set of vertices which have already been colored; for each vertex in $N(S)$ color it *white* or *black* such that the adjacent vertices have distinct colors, where $N(S)$ is the set of vertices which are adjacent to at least one vertex in $S$; repeat until $|S| = |V|$, where $|\ |$ denotes the cardinality of a set. In the end $V = C_1 \cup C_2$ with $C_1$, $C_2$ the *black* and *white* vertices, respectively.

Given $n$ objects to be clustered, we use a weighted and complete graph $G = (V, E)$ to represent the problem at hand, where $V$ denotes the set of objects ($|V| = n$, and hereinafter $n$ is the number of vertices of the input graph), and the weight $d(u, v)$ associated with an edge $(u, v) \in E$ denotes the dissimilarity between objects $u$ and $v$. Let $\wp$ denote the set of all partitions of $n$ objects into $k$ non-empty and disjoint clusters $\{C_1, C_2, \ldots, C_k\}$. For an object $p$ and a set $S$ of objects, we use $dMin(p, S)$ to denote the minimum dissimilarity between $p$ and objects in $S$,

$$dMin(p, S) = \min_{q \in S} d(p, q). \qquad (1)$$

**Definition 2.1.** For a partition $P = \{C_1, C_2, \ldots, C_k\} \in \wp$, the *split* $s(C_i)$ of $C_i$ is defined as (2), and the *split* $s(P)$ of $P$ is the minimum $s(C_i)$ among $i = 1, 2, \ldots, k$.

$$s(C_i) = \min_{p \in C_i, q \in C_j, j \in \{1,2,\cdots,k\}, j \neq i} d(p, q). \qquad (2)$$

**Definition 2.2.** For a cluster (or a set of objects) $C$, the *diameter* $d(C)$ *of* $C$ is defined as (3), and for a partition $P = \{C_1, C_2, \ldots, C_k\} \in \wp$, the *diameter* $d(P)$ of $P$ is the maximum diameter $d(C_i)$ of $C_i$ among $i = 1, 2, \ldots, k$.

$$d(C) = \max_{p,q \in C} d(p, q). \qquad (3)$$

**Definition 2.3.** The problem of *maximizing the ratio of split to diameter*, abbr. MRSD, is defined as (4),

$$\max_{P \in \wp} \frac{s(P)}{d(P)}. \qquad (4)$$

For any $P \in \wp$, the larger the ratio $s(P) / d(P)$ is, the more natural the partition will be. Specially, if $s(P) / d(P) \geq 1$, the dissimilarity between a pair of objects in the same cluster is always smaller than the dissimilarity between a pair of objects in different clusters, "*this is a strong property and means an excellent partition has been found when it holds*." [Delattre & Hansen, 1980].

## 3. The NP-hardness of MRSD

In this subsection, we assume that the weights associated with the edges of the input graph $G = (V, E)$ satisfy the triangle inequality, i.e., $\forall u, v, w \in V, d(u, v) \leq d(u, w) + d(v, w)$.

**Lemma 3.1** [Delattre & Hansen, 1980]**.** Given a weighted and connected graph $G = (V, E)$ (*complete* or *incomplete*), the number of distinct values of split among partitions $P \in \wp$ is at most $|V| - 1$. These values are equal to the weights of the edges of any minimum spanning tree of $G$.

**Definition 3.1.** A weighted and complete graph $G$ is *restricted* if $G$ has only one split value, i.e., the weights associated with edges of any minimum spanning tree of $G$ are equal to each other.

Consider the following decision problem (abbr. *the k-restricted MRSD*): given a restricted graph $G = (V, E)$, a positive integer $k$, and a positive value $\lambda$, does there exist a partition $P$ of $V$ into $k$ clusters such that $s(P) / d(P) \geq \lambda$? Clearly, the $k$-restricted MRSD $\in$ NP.

For $k = 3$, we prove the NP-completeness of the 3-restricted MRSD problem by reducing the 3-colorability problem to it. Recall that the 3-colorability problem is: given an unweighted graph $G = (V, E)$, is $G$ 3-colorable? Given any 3-colorability instance $G = (V, E)$, we construct a 3-restricted MRSD instance as follows: $G_c$ is a weighted and complete graph with the same vertices as $G$, i.e., $G_c =$ $(V, E \cup E^-)$, where $E^-$ is the set of edges of the complementary graph $G^-$ of $G$, and edges of $E$ are assigned a weight of 1 and those of $E^-$ are assigned a weight of 0.5. Clearly, the weights associated with the edges of $G_c$ satisfy the triangle inequality and the construction of $G_c$ can be done in polynomial time.

Now, we give the relations between the number of components of $G^-$ ($G^-$ may be disconnected) and the chromatic number $\chi(G)$ of $G$.

**Lemma 3.2.** Let $N$ be the number of components of $G^-$, then:

(a). If $N > 3$, then $\chi(G) \geq 4$.

(b). If $N = 3$ and all components are cliques, $\chi(G) \leq 3$; If $N = 3$ and at least one component is not a clique, $\chi(G) \geq 4$.

(c). If $N = 2$ and none of two components $C_1$ and $C_2$ is a clique, $\chi(G) \geq 4$; If $N = 2$ and all components are cliques, $\chi(G) = 2$; If $N = 2$ and only one component, e.g., $C_1$ is not a clique, then $G$ is 3-colorable if and only if $C_1$ is 2-colorable, therefore in this case, the 3-colorability problem for $G$ can be solved using the bicoloring algorithm for $C_1$.

**Proof.** (a). Arbitrarily select a vertex from each component and we get $N$ vertices. Since there does not exist an edge between any pair of those $N$ vertices in $G^-$, there must exist a clique with $N$ vertices in $G$, and the chromatic number of a clique with $N$ vertices is $N$.

(b). The *first* part: if all components are cliques, we can color the vertices of $G$ in the following way: all vertices of the $i$th component receive the color $i$ for $i = 1, 2, 3$. Since each component is a clique in $G^-$, any pair of vertices within the same component must not be adjacent to each other in $G$, hence the above coloring is feasible. The *second* part: since there is at least one component $C$ which is not a clique, there are at least two vertices $u$ and $v \in C$ such that $(u, v) \notin E^-$. These two vertices form a clique with four vertices in $G$ together with two other vertices arbitrarily selected from two other components respectively. Again, the chromatic number of a clique with four vertices is four.

(c). The *first* part: since none of components $C_1$ and $C_2$ is a clique, there are at least two vertices $u_1$ and $u_2 \in C_1$ such that $(u_1, u_2) \notin E^-$, and at least two vertices $v_1$ and $v_2 \in C_2$ such that $(v_1, v_2) \notin E^-$, these four vertices form a clique in $G$, and thus $\chi(G) \geq 4$. The *second* part: if all components are cliques, clearly $G$ can be colored in two colors according to the argument for the first part of (b), and thus $\chi(G) = 2$. The *third* part: if $C_1$ is 2-colorable, we can color all vertices of $C_2$ using another color not used by vertices of $C_1$, so $G$ is 3-colorable. If $G$ is 3-colorable, then $C_1$ must be 2-colorable, otherwise $\chi(G) \geq 4$ since any vertex $u$ of $C_2$ cannot be colored using the same color used by any vertex $v \in C_1$. □

Therefore, if $N \geq 2$, the problem of whether $G$ is 3-colorable can be solved according to the lemma 3.2. Now, we consider the case $N = 1$, i.e., $G^-$ is connected.

**Lemma 3.3.** If $G^-$ is connected, $G_c$ has only one split value, i.e., 0.5.

**Proof.** Since $G^-$ is a connected graph associated with weights of 0.5, $G^-$ must have a minimum spanning tree $T$ associated with weights of 0.5. Since $G^-$ is a subgraph connected all vertices of $G_c$ and any edge $(u, v)$ of $G_c$ such that $(u, v) \notin E^-$ has a weight of 1, $T$ must also be a minimum spanning tree of $G_c$, and hence $G_c$ has only one split value, i.e., 0.5, according to the lemma 3.1. □

**Lemma 3.4.** If $G^-$ is connected, then $G$ is 3-colorable if and only if there is a partition $P$ of $V$ into three clusters such that $s(P) / d(P) \geq 1$.

**Proof.** Since $G^-$ is connected, for any partition $P$ of $V$, $s(P) = 0.5$ by the lemma 3.3.

The *if* direction: since $s(P) / d(P) \geq 1$, we have $d(P) \leq 0.5$. Assume $P = \{C_1, C_2, C_3\}$, color all vertices of $C_i$ the color $i$ for $i = 1, 2, 3$. Since $d(C_i) \leq 0.5$ for $i = 1, 2, 3$, for any pair $u$ and $v$ of vertices of $C_i$, $(u, v) \notin E$, so the colorability is feasible.

The *only if* direction: since $G$ is 3-colorable, $V$ can be partitioned into three groups $C_1, C_2$, and $C_3$, such that $\forall u, v \in C_i$ for $i = 1, 2, 3$, $(u, v) \notin E$. Therefore, $\forall u, v \in C_i$ for $i = 1, 2, 3$, $(u, v) \in E^-$, equivalently $d(u, v) = 0.5$, which means that $P = \{C_1, C_2, C_3\}$ is a partition satisfying $s(P) / d(P) \geq 1$. □

Combining the lemma 3.2 with the lemma 3.4, we have the following lemma:

**Lemma 3.5.** The 3-restricted MRSD problem is NP-complete.

**Theorem 3.1.** For $k \geq 3$, the decision problem of MRSD is NP-complete.

**Proof.** Clearly, the decision problem of MRSD belongs to NP. We prove this theorem by reducing the 3-restricted MRSD to it. Given any 3-restricted MRSD instance $G = (V, E)$, let $d_{max}$ be the maximum weight in $E$, then we construct a MRSD instance as follows: $G' = (V', E')$, where $V' = V \cup A$, and $A = \{v_1, v_2, \ldots, v_{k-3}\}$ ($v_i$ is not used in $V$, $i = 1, 2, \ldots, k-3$); for edges $(u, v_i)$ and $(v_i, v_j)$ (where $u \in V$, $i, j = 1, 2, \ldots, k-3$), $d(u, v_i) = d(v_i, v_j) = d_{max} + 1$. Clearly, the weights associated with the edges of $G'$ satisfy the triangle inequality and the construction of $G'$ can be done in polynomial time. We prove that there is a partition $P = \{C_1, C_2, C_3\}$ of $V$ into three clusters such that $s(P) / d(P) \geq \lambda$ if and only if there is a partition $P'$ of $V'$ into $k$ clusters such that $s(P') / d(P') \geq \lambda$, and thus the theorem holds by the lemma 3.5. Let $s$ be the (unique) split of $G$, $P'$ be any partition of $V'$ such that at least a pair of vertices $u \in V$ and $v \in A$ are grouped into the same cluster, and $P''$ be any partition of $V'$ such that each vertex of $A$ forms a singleton cluster. Then according to the construction of $G'$, $s(P') = s(P'') = s$, and $d(P') = d_{max} + 1$, $d(P'') \leq d_{max}$, and hence $s(P') / d(P') \leq s(P'') / d(P'')$. Therefore, if there is a partition $P = \{C_1, C_2, C_3\}$ of $V$ into three clusters such that $s(P) / d(P) \geq \lambda$, then the partition $P'' = \{C_1, C_2, C_3, \{v_1\}, \{v_2\}, \ldots, \{v_{k-3}\}\}$ is a partition of $V'$ into $k$ clusters such that $s(P'') / d(P'') \geq \lambda$; if there is a partition $P' = \{C_1, C_2, \ldots, C_k\}$ of $V'$ into $k$ clusters such that $s(P') / d(P') \geq \lambda$, then there is a partition $P'' = \{C_1', C_2', C_3', \{v_1\}, \{v_2\}, \ldots, \{v_{k-3}\}\}$ such that the clusters $C_1'$, $C_2'$ and $C_3'$ consist of vertices of $V$ and $s(P'') / d(P'') \geq s(P') / d(P') \geq \lambda$, and hence the partition $P = \{C_1', C_2', C_3'\}$ is a partition of $V$ into three clusters such that $s(P) / d(P) \geq \lambda$. □

## 4. The Proposed Algorithms

We propose our algorithms based on the following observation:

**Theorem 4.1.** Given the input graph $G$, for any $v > 0$, the optimal MRSD solution $P^*$ of $G$ can be obtained by taking the larger of the maximum solution $P$ of $G$ with $s(P) \leq v$ and the maximum solution $P'$ of $G$ with $s(P') > v$.

**Proof.** The optimal solution $P^*$ of $G$ either separates the two vertices of some edge $e$ with $d(e) \leq v$ into different clusters, and thus $s(P^*) \leq v$; or $P^*$ does not separate the two vertices of any edge $e$ of $G$ with $d(e) \leq v$ into different clusters, equivalently, the two vertices of $e$ must be within the same cluster of $P^*$, and hence $s(P^*) > v$, then the maximum solution $P'$ of $G$ with $s(P') > v$ must be the optimal solution of $G$, and the theorem holds. □

Therefore, a recursion procedure on the split value can be used to obtain the proposed algorithms. Before presenting our algorithms, we first introduce the following concept.

**Definition 4.1.** Given a graph $G = (V, E)$, two vertices $u$ and $v$ adjacent to an edge $(u, v)$ are *merged*, or simply an edge $(u, v)$ is *merged*, means that the two vertices are replaced by a new vertex, the two edges from $u$ and $v$ to a remaining vertex are replaced by an edge weighted by the largest of the weights of the previous two edges and the edge $(u, v)$ is removed, and other edges together with their weights remain unchanged. We also call the new vertex a *supervertex* and $u$, $v$ the *merged vertices*.

It is noted that for a graph $G'$ obtained by merging the two vertices of some edge of another graph $G$, if the weights associated with the edges of $G$ satisfy the triangle inequality, then the weights associated with the edges of $G'$ also satisfy the triangle inequality. The proof is simple, and here we omit it considering the space limitation.

Let $G'$ be a graph obtained by merging the vertices of some edges of the input graph $G$, and $P'$ a partition of the vertices of $G'$. $P$ is the partition of vertices of $G$ induced by $P'$ means that $P$ is obtained from $P'$ by substituting $\{u_1,$ $u_2, \ldots, u_r\}$ for each supervertex $u$ of $P'$ (assume that $u$ is the supervertex by merging the vertices $u_1, u_2, \ldots, u_r$ of $G$, and denoted as $u = \{u_1, u_2, \ldots, u_r\}$). It is easy to verify the correctness of the following relation between $d(P)$ and $d(P')$: $d(P) = \max\{d(P'), d(u)\}$, where $d(u)$ is the maximum diameter among all supervertices of $P'$. According to the definition 4.1, if $u$ is a supervertex of $G'$, then for any other vertex $p$ of $G'$, the information about the minimum dissimilarity between $p$ and $u$ is lost in $G'$. So, the split $s(P')$ of $P'$ is meaningless and not used in the rest of the paper.

### 4.1 An Exact Algorithm for Bipartition

Since the algorithm uses the concept of *maximum spanning tree*, we first review it. Just as its name implies, a maximum spanning tree of a weighted graph $G$ is a spanning tree with the largest sum of weights associated with its edges among all spanning trees of $G$. A maximum spanning tree can be obtained using an algorithm similar to any minimum spanning tree algorithm, e.g., Kruskal's algorithm [Kruskal, 1956], and the only difference is that edges are considered in *descending* order of weights while in *ascending* order for seeking a minimum spanning tree.

The following lemma guarantees that a minimum diameter partition can be obtained for $k = 2$.

**Lemma 4.1.** For a weighted and connected graph $G$, the partition $P$ obtained by applying the bicoloring algorithm to a maximum spanning tree $T_{max}$ of $G$ has the minimum diameter.

**Proof.** According to the bicoloring algorithm, any pair of vertices $u$ and $v$ adjacent to an edge $(u, v)$ of $T_{max}$ must be in different clusters. So, for any edge $e = (u, v)$ of $G$, if the addition of it to $T_{max}$ will close an even cycle together with the unique path joining $u$ and $v$ in $T_{max}$, $u$ and $v$ must be in different clusters. Hence, the diameter of $P$ must be the largest edge $(p, q) \notin T_{max}$ such that it closes an odd cycle $C$ together with the unique path joining $p$ and $q$ in $T_{max}$, i.e., $s(P) = d(p, q)$. Let $P^* = \{C_1, C_2\}$ be the bipartition with the minimum diameter. We show that $d(P^*) \geq d(p, q)$ and hence $P$ is also a partition with the minimum diameter. By construction of $T_{max}$, $d(u, v) \geq d(p, q)$ for all edges $(u, v)$ of $C$ different from $(p, q)$, and hence, if $d(P^*) < d(p, q)$, any two vertices adjacent to an edge of $C$ should not belong to the same cluster. So, vertices in $C$ should alternatively belong to $C_1$ and $C_2$. But it is impossible since $C$ is odd. □

Now, we present an exact algorithm for $k = 2$ as shown in Fig. 1. Given the input graph $G$, the algorithm first constructs a minimum spanning tree $T_{min}$ of $G$, let $L$ be the list of edges of $T_{min}$ in ascending order of weights, and then repeats the following procedure until $L$ becomes empty: construct a maximum spanning tree $T_{max}$ of $G$; obtain the bipartition $P$ of vertices of $G$ by applying the bicoloring algorithm in Section 2 to $T_{max}$; if the cost of the current bipartition is larger than the previous cost, then

**Clustering to Maximize the Ratio of Split to Diameter**

```
Algorithm: MRSD_Bipartition
Input: a weighted and complete graph G = (V, E).
Output: the bipartition solution of V.

Cost ← 0;      /*the cost of the optimal solution*/
solution ← Φ;  /* the optimal solution*/
Construct a minimum spanning tree T_min of G;
L ← The edges of T_min in ascending order of weights;
G' = (V', E') ← G;              /*a copy of the input graph*/
while(|L| ≥ 1)
    Construct a maximum spanning tree T_max of G';
    Let P' be the bipartition of V' obtained by applying the
    bicoloring algorithm to T_max;
    Let P be the partition of V induced by P';
    if(s(P) / d(P) > Cost)
        Cost ← s(P) / d(P);
        solution ← P;
    end if
    (p, q) ← the first edge in L;
    while(d(p, q) ≤ s(P))
        Find the vertices S_1, S_2 of G' such that p is merged into S_1
        (or p = S_1) and q is merged into S_2 (or q = S_2) respectively;
        Merge the two vertices S_1 and S_2 of G';
        L ← L − {(p, q)};
        if(|L| ≥ 1)
            (p, q) ← the first edge in L;
        else
            break;
        end if
    end while
end while
return solution;
```

*Figure 1.* An exact algorithm for $k = 2$.

update the solution; merge the two vertices of edges $(p, q)$ of $L$ such that $d(p, q) \le s(P)$.

**Lemma 4.2.** Let $P_i$ be the partition of vertices of the input graph $G$ found by *MRSD_Bipartition* (or *MRSD_Multipartition* in the next subsection), $i = 1, 2, …, r$, then $s(P_i)$ is strictly monotone increasing, i.e., $s(P_1) < s(P_2) < s(P_3) < … < s(P_r)$.

**Proof.** First, we show that if the edges $e_T$ of the minimum spanning tree $T_{min}$ of $G$ with $d(e_T) \le s(P_i)$, $i = 1, 2, …, r$, have been merged, then the two vertices of any edge $e_G$ of $G$ with $d(e_G) \le s(P_i)$ must be within the same supervertex: if $e_G \in T_{min}$, it is evident; if $e_G = (u, v) \notin T_{min}$, then $d(u, v) \ge d(e)$ for all edges $e$ in the path $X$ from $u$ to $v$ in $T_{min}$ since if $d(u, v) < d(e)$ for some edge $e$, then we can get a lighter spanning tree $T'$ of $G$ than $T_{min}$ by adding the edge $(u, v)$ into $T_{min}$ and deleting the edge $e$; so, if $d(u, v) \le s(P_i)$, we have $d(e) \le d(u, v) \le s(P_i)$ for any $e \in X$, and hence all vertices in $X$, of course including $u$ and $v$, must be within the same supervertex. Therefore, $s(P_i) < s(P_{i+1})$ for $i = 1, 2, …, r – 1$ since the edges $e$ in $G$ with $d(e) \le s(P_i)$ have been merged before the partition $P_{i+1}$ is found. □

**Lemma 4.3.** Let $G$ be the input graph and $P$ the current partition of vertices of $G$ found by *MRSD_Bipartition*. Then after the solution is updated if necessary, the relation $Cost(solution) \ge Cost(Q)$ holds for all partitions $Q$ of vertices of $G$ with split value $s(Q) \le s(P)$.

**Proof.** We use the mathematical induction on the index $i$ of partitions $P_i$ of vertices of $G$ for $i = 1, 2, …, r$.

*The basic step $i = 1$.* Then $G' = G$ and $solution = P_1$. For any partition $Q$ of vertices of $G$ with split $s(Q) \le s(P_1)$, we have $d(Q) \ge d(P_1)$ by the lemma 4.1, and thus $Cost(Q) = s(Q) / d(Q) \le s(P_1) / d(P_1) = Cost(solution)$ holds.

*The induction hypothesis.* Assume that the lemma holds for $i = 1, 2, …, l$ $(l < r)$.

Now, we demonstrate that the lemma also holds for $i = l + 1$. According to the induction hypothesis, we have $Cost(solution) \ge Cost(Q)$ for all partitions $Q$ of vertices of $G$ with $s(Q) \le s(P_l)$. By the lemma 4.2, we have $s(P_l) < s(P_{l+1})$, so we only need to show that $Cost(solution) \ge Cost(Q)$ for all partitions $Q$ with $s(P_l) < s(Q) \le s(P_{l+1})$. Note that $d(P_{l+1}) = \max\{d(P'_{l+1}), d(u)\}$ and $d(Q) = \max\{d(Q'), d(u)\}$, where $P'_{l+1}$ is the partition of vertices of $G'$ by which the partition $P_{l+1}$ is induced, $d(u)$ is the maximum diameter among all supervertices of $G'$, and $Q'$ is any partition of vertices of $G'$ by which the partition $Q$ is induced. Since $d(P'_{l+1}) \le d(Q')$ by the lemma 4.1, we have $d(P_{l+1}) \le d(Q)$, and hence $Cost(Q) = s(Q) / d(Q) \le s(P_{l+1}) / d(P_{l+1}) = Cost(P_{l+1})$. If $Cost(P_{l+1}) > Cost(solution)$, then *solution* is replaced by $P_{l+1}$ according to the algorithm, and the lemma holds; if $Cost(P_{l+1}) \le Cost(solution)$, the lemma trivially holds. □

**Lemma 4.4.** Let $s^*$ be the largest split value of the input graph $G$, and $P$ the last partition of vertices of $G$ found by the algorithm *MRSD_Bipartition* (or *MRSD_Multipartition* in the next subsection), then $s(P) = s^*$.

**Proof.** Note that $G'$ has $m$ vertices if and only if $(m − 1)$ edges remain in $L$ $(1 \le m \le n)$. According to the definition of split, the vertices of $G$ can be grouped into $k$ disjoint and nonempty subsets $C_1, C_2, …, C_k$ such that for any pair of vertices $u \in C_i$ and $v \in C_j$ $(i, j = 1, 2, …, k$, and $i \ne j)$, we have $d(u, v) \ge s^*$ and there exists at least one edge $(u, v)$ such that $d(u, v) = s^*$.

Therefore if $s(P) < s^*$, then after merging edges $e$ of $T_{min}$ with $d(e) \le s(P) < s^*$, there are at least $k$ vertices in the current $G'$ (or equivalently $(k − 1)$ edges remain in $L$) since any pair of vertices $u \in C_i$ and $v \in C_j$ $(i, j = 1, 2, …, k$, and $i \ne j)$ have not been merged, and hence $P$ can not be the last partition of vertices of $G$ found by the algorithm *MRSD_Bipartition* (or *MRSD_Multipartition*). □

Combining the lemma 4.3 with the lemma 4.4, we have the following theorem:

**Theorem 4.2.** The algorithm *MRSD_Bipartition* finds the optimal solution of MRSD for $k = 2$.

**Theorem 4.3.** The worst-case runtime of *MRSD_Bipartition* is $O(n^3)$.

**Proof.** Let $V'$ be the set of vertices of $G'$. In each iteration, it needs time $O(|V'|^2)$ to find a maximum spanning tree $T_{max}$ of $G'$ [Cormen et al, 2001], $O(|V'|)$ to bicolor the vertices of $T_{max}$, and $O(|V'|^2)$ to compute the diameter and

```
Algorithm: MRSD_Multipartition
Input: a weighted and complete graph G = (V, E), and the
       number k of clusters.
Output: the partition solution of V.

Cost ← 0;          /*the cost of the solution*/
solution ← ∅;      /*an approximation solution*/
Construct a minimum spanning tree T_min of G;
L ← The edges of T_min in ascending order of weights;
G' = (V', E') ← G;  /*a copy of the input graph*/
while(|L| ≥ k − 1)
   S ← ∅;
   Randomly select a vertex p from V';
   S ← S ∪ {p};
   while(|S| < k)     /*select k representatives*/
      p ← arg max dMin(q, S);
            q∉S
      S ← S ∪ {p};
   end while
   Assign each vertex p of V' to its nearest representative in S
       and assume that the partition of V' is P';
   Let P be the partition of V induced by P';
   if(s(P) / d(P) > Cost)
      Cost ← s(P) / d(P);
      solution ← P;
   end if
   (p, q) ← the first element in L;
   while(d(p, q) ≤ s(P))
      Find the vertices S_1, S_2 of G' such that p is merged into S_1
         (or p = S_1) and q is merged into S_2 respectively (or q = S_2);
      Merge the two vertices S_1 and S_2 of G';
      L ← L − {(p, q)};
      if(|L| ≥ k − 1)
         (p, q) ← the first edge in L;
      else
         break;
      end if
   end while
end while
return solution;
```

*Figure 2.* An approximation algorithm for $k \geq 3$.

split of a partition. Since $|L| = n − 1$, the loop body executes at most $n − 1$ times, therefore the overall steps for the above three procedures are $O(n^3)$. As to the merging operations, we can use the disjoint-set forest data structure with *union by rank* and *path compression* [Cormen et al, 2001]: initially, use $n$ MAKE-SET operations to construct $n$ trees and each tree consist of one vertex of the input graph; use FIND-SET operation to find in which tree the merged vertex is, and use UNION operation to merge two trees. Note that a sequence of $m$ MAKE-SET, UNION, and FIND-SET operations can be performed on a disjoint-set forest in time $O(m)$ in all practical situations. Since $|L| = n − 1$, the number of all merging operations is $O(n)$, and a merging operation takes time $O(|V|)$ to compute the maximum dissimilarities between other vertices and the two merged vertices, so all merging operations take time $O(n^2)$. The overall time of *MRSD_Bipartition* is $O(n^3) + O(n^2) = O(n^3)$. □

It is noted that we do not use the triangle inequality to prove the correctness of the algorithm in this subsection and the lemma 4.1, so the algorithm *MRSD_Bipartition* can be applied to the input graph even if the weights do not satisfy the triangle inequality.

### 4.2 An Approximation Algorithm for Multipartition

We present an approximation algorithm for $k \geq 3$ as shown in Fig. 2 on the precondition that the weights associated with edges of the input graph satisfy the triangle inequality.

Given the input graph $G$ and the number $k$ of clusters, the algorithm first constructs a minimum spanning tree $T_{min}$ of $G$, let $L$ be the list of edges of $T_{min}$ in ascending order of weights, and then repeats the following procedure until $|L| < k − 1$: use the farthest-point clustering [Gonzalez, 1985] to select $k$ representatives from vertices of $G'$, and obtain a partition $P'$ of vertices of $G'$ by assigning each vertex to its nearest representative; if the cost of the current partition $P$ induced by $P'$ is larger than the previous cost, then update the solution; merge the two vertices of edges $(p, q)$ of $L$ such that $d(p, q) \leq s(P)$. The $k$ representatives are selected as follows: first, randomly select a vertex as the first representative and add it into the set $S$ of representatives, then select the vertex $p$ into $S$ such that $dMin(p, S)$ is maximum among all unselected vertices, and repeat until $|S| = k$.

**Lemma 4.5.** Let $G$ be the input graph and $P$ the partition of vertices of $G$ induced by the partition of vertices of the current $G'$ found by *MRSD_Multipartition*. If the split value $s(P^*)$ of the optimal solution $P^*$ satisfies $s(P^*) \leq s(P)$, then after the solution is updated if necessary, $Cost(solution) \geq Cost(P^*) / 2$.

**Proof.** We use the mathematical induction on the index $i$ of partitions $P_i$ of vertices of $G$ for $i = 1, 2, …, r$.

*The basic step* $i = 1$. Then $G' = G$ and $solution = P_1$. Let $S$ be the set of selected $k$ representatives in the current iteration. Consider the vertex $p$ which maximizes the minimum dissimilarity between it and $S$, in other words, the object that would be chosen if we picked one more representative. Assume that $dMin(p, S) = \delta$, then all pairwise dissimilarities among $S \cup \{p\}$ are at least $\delta$. In any partition of vertices of $G'$ into $k$ clusters, at least two of these vertices must be within the same cluster, so $d(P^*) \geq \delta$. Since $d(P_1)$ is at most $2\delta$ by the triangle inequality, so if $s(P^*) \leq s(P_1)$, we have $Cost(solution) = s(P_1) / d(P_1) \geq s(P^*) / (2\delta) \geq s(P^*) / (2d(P^*)) = Cost(P^*) / 2$.

*The induction hypothesis.* Assume that the lemma holds for $i = 1, 2, …, l$ ($l < r$).

We demonstrate that the lemma also holds for $i = l + 1$. We have $Cost(solution) \geq Cost(P^*) / 2$ if $s(P^*) \leq s(P_l)$ by the induction hypothesis. By the lemma 4.2, we have $s(P_l) < s(P_{l+1})$, so we only need to show that $Cost(solution) \geq Cost(P^*) / 2$ if $s(P_l) < s(P^*) \leq s(P_{l+1})$. Note that $d(P_{l+1}) = \max\{d(P'_{l+1}), d(u)\}$ and $d(P^*) = \max\{d(P^{*\prime}), d(u)\}$, where $P'_{l+1}$ is the partition of vertices of $G'$ by which the partition $P_{l+1}$ is induced, $d(u)$ is the maximum diameter among all supervertices of $G'$, and $P^{*\prime}$ is the partition of vertices of $G'$ by which $P^*$ is induced. Let $S$ be the set of selected $k$ representatives from vertices of $G'$ and $p$ the vertex which maximizes the minimum dissimilarity between it and $S$. Assume that $dMin(p, S) = \delta$, then as the same arguments in the basic step, we have $d(P'_{l+1}) \leq 2\delta$ and $d(P^{*\prime}) \geq \delta$, and hence no matter what relations are between $d(P'_{l+1})$ and $d(u)$ and between $d(P^{*\prime})$ and $d(u)$, the relation $d(P_{l+1}) \leq 2d(P^*)$ always holds. Therefore, if $s(P_l) < s(P^*) \leq s(P_{l+1})$, then $Cost(P_{l+1}) = s(P_{l+1}) / d(P_{l+1}) \geq s(P^*) / 2d(P^*) = Cost(P^*) / 2$. If $Cost(P_{l+1}) > Cost(solution)$, then *solution* is replaced by $P_{l+1}$ according to the algorithm, and the lemma holds; if $Cost(P_{l+1}) \leq Cost(solution)$, then the lemma trivially holds. □

Combining the lemma 4.5 with the lemma 4.4, we have the following theorem:

**Theorem 4.4.** The algorithm *MRSD_Multipartition* is a factor of 2 approximation algorithm of MRSD for $k \geq 3$.

As the same arguments for Theorem 4.3, we have the following theorem:

**Theorem 4.5.** The worst-case runtime of *MRSD_Multipartition* is $O(n^3)$.

## 5. The Experimental Results

We evaluate the proposed algorithms by applying them to image segmentation and compare the proposed algorithms with the popular Normalized Cut (abbr. NCut) [Shi & Malik, 2000] in the computer vision field. At the same time, we also give the results of the complete-linkage algorithm (abbr. CLA) and the single-linkage algorithm (abbr. SLA) to verify whether the proposed criterion can overcome the *dissection* and *chain* effects which may be resulted in by CLA and SLA respectively. For image segmentation, the problem of how to define features by incorporating a variety of cues is a nontrivial one. In our experiments, we simply use the color features (the RGB model) and spatial proximity of pixels to compute the dissimilarity between a pair of pixels, since the focus of this paper is on developing a general clustering algorithm, given a dissimilarity measure.

We use Manhattan distance $d(i, j) = d_1(i, j) + d_2(i, j)$ to compute the dissimilarity between a pair of pixels $i$ and $j$, where $I_c(i)$ is the *red*, *green*, or *blue* color component for pixel $i$, $row(i)$ is the row number of pixel $i$ and $col(i)$ is the column number of pixel $i$

$$\begin{cases} d_1(i,j) = \sum_{c \in \{R,G,B\}} |I_c(i) - I_c(j)| \\ d_2(i,j) = |row(i) - row(j)| + |col(i) - col(j)| \end{cases}, \quad (5)$$

Since NCut uses a similarity matrix as the input, we use (6) suggested in [Shi & Malik, 2000] to compute the similarity $w(i, j)$ between a pair of pixels $i$ and $j$,

$$w(i,j) = e^{-(d_1(i,j)/\sigma_I)^2} \times \begin{cases} e^{-(d_2(i,j)/\sigma_X)^2} & \text{if } d_2(i,j) < r \\ 0 & \text{else} \end{cases}, \quad (6)$$

where $\sigma_I$ and $\sigma_X$ are two parameters and are typically set to 10 to 20 percent of the ranges of the color feature dissimilarity $d_1$ and the spatial proximity dissimilarity $d_2$ respectively suggested in [Shi & Malik, 2000], and $r$ is a parameter for constructing a locally connected (or sparse) graph to accelerate the computing of eigenvalues involved in NCut. Note that if $d_2(i, j) \geq r$ for any pair of pixels $i$ and $j$, then the similarity $w(i, j) = 0$.

We conduct the experiments on two groups of images: seven natural images and two synthetic images, which are used to test whether the proposed algorithms and NCut can find the desired clusters. The size of all images are 60×60 pixels. All algorithms are implemented in MATLAB R2009b, and experiments are carried out on a 2.6 GHz Pentium Dual-Core with 2G bytes of RAM. For NCut, there are two experiment setups named *NCut_C* and *NCut_S* respectively. *NCut_C*: $\sigma_I$ and $\sigma_X$ are set to 0.15; $r = +\infty$, i.e., the input similarity graph is complete; and the function used to solve the eigenvalues is the MATLAB function *eig*, which computes all eigenvalues of a dense matrix. *NCut_S*: $\sigma_I$ and $\sigma_X$ are set to 0.15; $r = 5$, i.e., the input similarity graph is locally connected and thus sparse; and the function used to solve the eigenvalues is the MATLAB function *eigs*, which computes at most six largest or smallest magnitude eigenvalues of a sparse matrix. Here, we compute the *two smallest* eigenvalues since NCut uses the eigenvector corresponding to the second smallest eigenvlaue to segment an image.

Fig. 3 depicts the clustering results of seven natural images and Fig. 4 depicts the clustering results of two synthetic images. For two synthetic images, the pixels with *white color* are to be clustered. Table 1 summarizes the runtime of the proposed algorithms and NCut on seven natural images.

For natural images, we observe that the clustering results obtained by *MRSD_Bipartition* and *NCut_C* are almost the same except the second image, on which the result of *MRSD_Bipartition* is a little better. However, the runtime of *NCut_C* is almost ten times of that of *MRSD_Bipartition*, although the worst-case time complexity of both algorithms is $O(n^3)$. The reason may be that *NCut_C* needs to solve the eigenvalues of a dense $l \times l$ matrix, where $l$ is the number of pixels (here $l = 3600$), and thus the upper bound of time complexity of *NCut_C* is much tighter than that of *MRSD_Bipartition*. For sparse graphs, although the runtime of *NCut_S* decreases about 20 percent than that of *NCut_C*, the results are poor.

For synthetic images, we observe that only the proposed algorithms obtain the desired results on two images: *NCut_C* does not get the desired results on any image, and *NCut_S* gets the desired result just on one image.

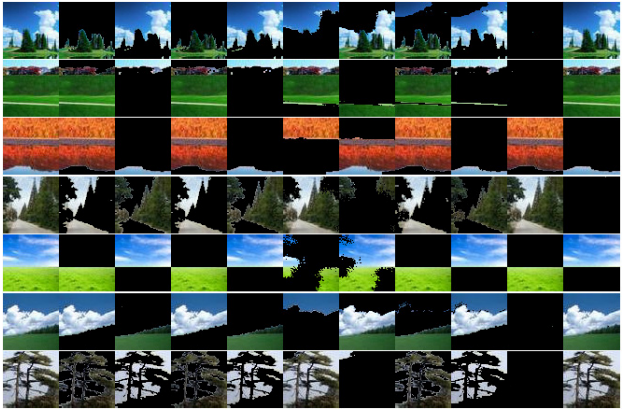

Figure 3. k = 2. a: the images to be clustered. b and c: the results of MRSD_Bipartition. d and e: the results of NCut_C. f and g: the results of NCut_S. h and i: the results of CLA. j and k: the results of SLA.

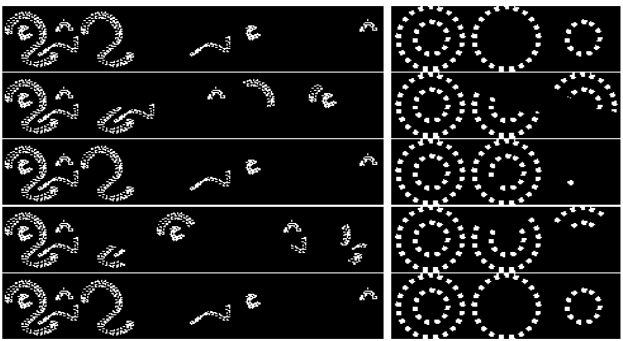

Figure 4. The left part: k = 4. The right part: k = 2. The images in the first column of two parts are to be clustered. The *first* row: the results of MRSD_Multipartition (left) or MRSD_Bipartition (right). The *second* row: the results of NCut_C. The *third* row: the results of NCut_S. The *fourth* row: the results of CLA. The *fifth* row: the results of SLA. Since NCut is a bipartition algorithm, the results for k = 4 are obtained by recursive call it.

Table 1. The runtime on seven natural images (seconds).

| No | MRSD | NCut_C | NCut_S |
|---|---|---|---|
| 1 | 91.322 | 1023.547 | 797.078 |
| 2 | 89.547 | 1003.797 | 789.188 |
| 3 | 124.391 | 1010.891 | 791.891 |
| 4 | 95.547 | 1022.875 | 809.281 |
| 5 | 88.547 | 1011.297 | 807.516 |
| 6 | 101.531 | 1011.766 | 789.891 |
| 7 | 126.359 | 1028.891 | 799.500 |
| Average | 102.463 | 1016.152 | 797.763 |

The results of CLA on the first, the second, and the sixth natural images are worse than that of the proposed algorithm, and CLA does not get the desired result on any synthetic image; the results of SLA on all natural images except the third and the fifth are worse than that of the proposed algorithm (note that some results just contain outliers), but SLA also gets the desired results on two synthetic images. The results demonstrate that the proposed criterion can partly overcome the *dissection* and *chain* effects resulted in by CLA and SLA respectively.

## 6. Conclusion

In this paper, we have proposed a new clustering criterion: maximization the ratio of the minimum split to the maximum diameter. This clustering criterion seems very natural and captures both the homogeneity and the separation conditions. For k = 2, an exact algorithm is presented, and for k ≥ 3, the problem is proven to be NP-hard and a factor of 2 approximation algorithm is presented if the weights associated with the input graph satisfy the triangle inequality. The experimental results on natural and synthetic images demonstrate the effectiveness of the proposed algorithms.

## Acknowledgments

This work was supported by China Natural Science Foundation under Grant No 60973083, Natural Science Foundation of Guangdong Province under Grant No 06300170, and the Fundamental Research Fund for the Central Universities, SCUT, under Grant No 2009ZM0175.